\documentclass[12pt,tightenlines]{revtex4}
\usepackage{amsfonts,}
\usepackage{amsmath}
\usepackage{txfonts}
\usepackage{graphicx}
\usepackage{amssymb}
\usepackage{color}
\usepackage{wrapfig,lipsum,booktabs}
\usepackage{floatrow}
\usepackage{epstopdf}
\usepackage{hyperref}
\usepackage[normalem]{ulem}

\begin{document}
\title{From topological analyses to functional modeling: the case of hippocampus} 
\author{Yuri Dabaghian\textsuperscript}
\affiliation{Department of Neurology, The University of Texas McGovern Medical School, 6431 Fannin St, Houston, TX 77030\\
$^{*}$e-mail: dabaghian@gmail.com}
\vspace{17 mm}
\date{\today}
\vspace{17 mm}

\begin{abstract}
Topological data analyses are rapidly turning into key tools for quantifying large volumes of neurobiological 
data, e.g., for organizing the spiking outputs of large neuronal ensembles and thus gaining insights into the 
information produced by various networks. Below we discuss a case in which several convergent topological analyses 
not only provide a description of the data structure, but also produce insights into how these data may be 
processed in the hippocampus---a brain part that plays a key role in learning and memory. 
The resulting functional model provides a unifying framework for integrating spiking information at different 
timescales and understanding the course of spatial learning at different levels of spatiotemporal granularity.
In particular, the model allows quantifying contributions of various physiological phenomena---brain waves, 
synaptic strengths, synaptic architectures, etc., into spatial cognition.

\end{abstract}
\maketitle

\newpage

\section{Introduction}
\label{section:intro}

Spatial cognition in mammals is based on an internal representation of their environments---a cognitive map---used 
for spatial planning, navigating paths, finding shortcuts, remembering location of the home nest, food sources and 
so forth. A central role in producing these maps is played by the hippocampal neurons famous for their spatially tuned
spiking activity. In rats, these neurons, known as ``place cells,'' fire in specific domains of the navigated 
environment---their respective ``place fields'' \cite{OKeefe,Moser}. Thus, the spatial layout of the place fields 
in a given environment $\mathcal{E}$---a place field map $M_{\mathcal{E}}$---defines the temporal order in which 
place cells fire during animal's moves \cite{Schmidt,Agarwal}, and is therefore viewed as a geometric ``proxy'' of 
the animal's cognitive map.

Experiments in `morphing' $2D$ environments demonstrate that place field maps are flexible: if the environment is 
deformed, the the place fields may change their shapes, sizes and locations, while preserving mutual overlaps, 
adjacencies, containments, etc. \cite{Gothard,Leutgeb,Wills,Touretzky,eLife}. Hence the sequences in which the 
place cells fire during animal's navigation remain largely invariant within a certain range of geometric transformations, 
which suggests that the hippocampus provides a qualitative, topological representation of space \cite{Poucet,Alvernhe,Wu,eLife}.

The mechanisms that produce cognitive maps and the computational principles by which the brain converts patterns 
of neuronal firing into global representations of external space remain vague. Broadly, it is believed that the 
information provided by the individual place cells is somehow combined into single coherent whole. However, this 
`fusion' should not be viewed as a na\"ive aggregation of the smaller `pieces,' because the signals provided by the 
individual neurons have no intrinsic spatial attributes; rather, spatial properties are \textit{emergent}, i.e., 
appearing at a neuronal ensemble level \cite{Wilson,Pouget,Postle}. 

A computational framework developed in \cite{PLoS,Arai,Basso,Hoffman,CAs,Efficacies,SchemaM,SchemaS,Novikov} helps 
to understand these phenomena by integrating the activity of the individual neurons into a large-scale map of the 
environment and to study the dynamics of its appearance, using algebraic topology techniques. Below we review some
basic ideas and key concepts used in this framework, and discuss how they may apply to hippocampal physiology and
cognitive realm. We then outline several examples that demonstrate how various characteristics of individual cells 
and synapses can be incorporated into the model and what effect these `microscopic' parameters produce at `macroscale,' 
i.e., in the map that they jointly encode. 

\section{Topological Model}
\label{section:model}

\textbf{Alexandrov-\v{C}ech's theorem}. The topological nature of the cognitive map suggests that the information 
transmitted via place cell spiking should be amenable to topological analyses. For example, a place field map can 
be viewed as a cover of the environment $\mathcal{E}$ by the place fields $\upsilon_i$, $\mathcal{E}=\cup_i\upsilon_i$,
which, according to the Alexandrov-\v{C}ech's theorem \cite{Alex,Cech}, encodes the topological shape of $\mathcal{E}$. 
Specifically, one constructs the nerve of the cover---an abstract simplicial complex $\mathcal{N}$ whose simplexes, 
$\nu_{i_0,i_1,\ldots,i_k}=[ \upsilon_{i_0},\upsilon_{i_1},\ldots \upsilon_{i_k}]$, correspond to nonempty overlaps 
between the place fields, $\upsilon_{i_0}\cap\upsilon_{i_1}\cap\ldots\cap\upsilon_{i_k}\neq \varnothing$. If these 
overlaps are contractible, then $\mathcal{N}$ has the same topological shape as $\mathcal{E}$, i.e., the same number 
of components, holes, tunnels, etc. \cite{Hatcher}. An implication of this construction is that if the place fields 
cover the environment sufficiently densely, then their overlaps encode its topology, which provides a link between 
the place cells' spiking pattern and the topology of the represented space \cite{Curto,Ghrist1,PLoS,Arai,Basso,Hoffman}.

\textbf{Temporal coactivity complex}. From the physiological perspective, the arguments based on the analyses of place
fields provide only an indirect description of the information processing in the brain. In reality, the hippocampus 
and the downstream brain regions do not have access to the shapes and the locations of the place fields, which are but 
artificial constructs used by experimentalists to visualize their data. In the brain, the information is transmitted
via neuronal spiking activity: if the animal enters a location where several place fields overlap, then there is 
a \textit{probability} that the corresponding place cells will produce spike trains that overlap \textit{temporally} 
\cite{Ghrist1,Curto,PLoS}. Such coactivities may be interpreted \textit{intrinsically} by the downstream brain areas, 
and integrated into a global map of the ambient space. Thus, a proper description of place cell (co)activity requires 
a \textit{temporal} analogue of the nerve complex built using temporal relationships between spike trains---which is, 
in fact, straightforward. Indeed, since the place field overlaps represent place cells' coactivities, one can construct 
a `coactivity complex' $\mathcal{T}$ whose simplexes correspond to combinations of active place cells, $\sigma = [c_{i_0}, 
c_{i_1},..., c_{i_k}]$. It was shown in \cite{Ghrist1,Curto,PLoS} that if such a complex is sufficiently large (i.e., 
if it incorporates a sufficient number of the coactivity events) then its structure is similar to the structure of the 
spatially-derived nerve complex $\mathcal{N}$, e.g., $\mathcal{T}$ correctly captures the topology of the physical 
environment.

\textbf{Simplicial schemas of cognitive maps}. Both complexes $\mathcal{N}$ and $\mathcal{T}$ provide a contextual 
framework for representing spatial information encoded by the place cells \cite{SchemaS}. For example, a sequence of 
place fields traversed during the rat's moves over a particular trajectory $\gamma$ and the place cell combinations 
ignited along this trajectory can be represented, respectively, by a `nerve path' $\Gamma_{\mathcal{N}}=\{\nu_1,\nu_2,
\ldots,\nu_k\}$---a chain of nerve-simplexes in $\mathcal{N}$, or by a `coactivity path' $\Gamma_{\mathcal{T}}=
\{\sigma_1,\sigma_2,\ldots,\sigma_k\}$---a sequence of the coactivity-simplexes in $\mathcal{T}$ (see also \cite{SchemaM}). 
These simplicial paths qualitatively represent the shape of the physical 
trajectories: a closed simplicial path represents a closed physical rout; a non-contractible simplicial path corresponds 
to a class of the physical paths that enclose unreachable or yet unexplored parts of the environment; two topologically 
equivalent simplicial paths $\Gamma_1 \sim \Gamma_2$ represent physical paths $\gamma_1$ and $\gamma_2$ that can be 
deformed into one another and so forth \cite{Brown1,Jensen1,Guger,Novikov}.

By the Alexandrov-\v{C}ech's theorem, the net pool of the simplicial paths can thus be used to describe the topological 
connectivity of the environment $\mathcal{E}$ via homological characteristics of $\mathcal{N}$ and $\mathcal{T}$. For 
example, the number of topologically inequivalent simplicial paths contracting to a vertex define the zeroth Betti 
number of the corresponding complexes, $b_0(\mathcal{N})$ or $b_0(\mathcal{T})$, and hence determine the number of 
disconnected components in $\mathcal{E}$, while the paths contracting to $1D$ loops define the first Betti numbers,
$b_1(\mathcal{N})$ or $b_1(\mathcal{T})$, that count holes in $\mathcal{E}$, etc. \cite{Hatcher}.

\textbf{A model of spatial learning}. A key difference between the complexes $\mathcal{N}$ and $\mathcal{T}$ is that
the topological shape of $\mathcal{N}$ is fully defined by the structure of the place field map, whereas the shape of 
$\mathcal{T}$ unfolds in time at the rate with which the spike trains are produced. At every given moment of time, the 
coactivity complex $\mathcal{T}$ represents connections between the place fields that the animal had time to ``probe:" as 
the animal begins to explore a new environment, $\mathcal{T}$ is small, fragmented and may contain gaps that represent
lacunae in the animal's internal map of the navigated space, rather than physical obstacles or inaccessible spatial 
domains. As the animal continues to navigate, more combinations of coactive place cells contribute connectivity information, 
the coactivity complex grows, $\mathcal{T}(t)\subseteq\mathcal{T}(t')$, $t<t'$, and acquires more details, converging to a 
stable shape that captures the physical structure of the surroundings.

Mathematically, $\mathcal{T}$ can thus be viewed as a \textit{filtered} complex, with the filtration defined by the 
times of the simplexes' first appearance, $t_{\sigma}$. Methods of the Persistent Homology theory allow describing 
the dynamics of the topological loops in $\mathcal{T}$, e.g., evaluating the minimal time $T_{\min}$ after which the 
topological structure of $\mathcal{T}$ matches the topology of the environment, $b_n(\mathcal{T}) = b_n(\mathcal{E})$
\cite{Zomorodian2,Ghrist2,Edelsbrunner}. Biologically, this value provides a low-bound theoretical estimate for the 
time required to learn a novel topological map from place cell outputs (Fig.\ref{Figure1}A) 
\cite{PLoS,Arai,Basso,Hoffman,CAs,Efficacies,SchemaM,SchemaS}.

\textbf{Facing the biological realm}. The physiological viability of these algebraic-topological constructions
depends on the parameters of neuronal firing activity: just as there must be a sufficient number of place fields 
covering a space in order to produce a topologically correct nerve complex $\mathcal{N}$, certain conditions must 
be met by the place cell spiking profiles in order to produce an operational coactivity complex $\mathcal{T}$. For 
example, there should be enough cofiring of place cells with sufficient spatial specificity of spiking; the encoded 
relationships should not be washed out by noise; the model should make realistic predictions, e.g., produce viable 
learning periods in different environments, etc. Given that biological systems are highly variable \cite{FentonVar}, 
these criteria may or may not be met by the physiological place cell ensembles, or vice versa, the model may single 
out a certain `operational' scope of parameters that may not match the biological range. Below we demonstrate that,
first, the model  can incorporate a vast scope of physiologically relevant characteristics of spike times, spiking 
statistics, their modulations by the `brain waves,' efficacies of synaptic connections, architectures of the neuronal 
networks, etc., all of which correlate with dynamics of spatial learning. Second, the model allows converting this 
information \textit{consistently} into coherent and biologically viable descriptions of a wide scope of neurophysiological 
phenomena. Third, it becomes possible to deduce functional properties of the system following not only empirical 
observations or experimental line of reasoning that currently dominate neurophysiological literature, but also the 
models' intrinsic logic.

\textbf{Parameterization}. To cope with the complexity of the cognitive map's construction, the model is built 
hierarchically: its main components implement most prominent physiological phenomena, and more subtle effects are 
incorporated as modifications of the skeletal structures. In the following, we will proceed in steps, by selecting 
a specific phenomenon, embedding it into the model using a minimal set of tools, describing the results and discussing 
biological implications.

To simplify the approach, we will describe neuronal spiking in terms of Poisson firing rates, which, in case of the 
place cells, can be approximated by Gaussian functions of rat's coordinates with the amplitude $f_i$ (the $i^{th}$ 
place cell's maximal firing rate), and the width $s_i$ (the size of the corresponding place field) \cite{Barbieri2,PLoS}. 
For an ensemble of $N$ place cells, the $N$ values $s_i$ and $f_i$ can be viewed as instantiations of two random 
variables drawn from their respective distributions with certain modes ($s$ and $f$ correspondingly) and standard 
deviations, $\sigma_s$ and $\sigma_f$. To avoid overly broad or overly narrow distributions we impose additional 
conditions $\sigma_s = bs$ and $\sigma_f = af$ with the coefficients $a$ and $b$ selected so match the experimental 
statistics \cite{BuzLog,Barbour,Brunel}. As a result, each specific place cell ensemble can be indexed by a triplet 
of parameters, $(s,f,N)$.

Second block of parameters characterizes animal's behavior, e.g., speeds and trajectories, which are computationally
intractable. We assume a practical approach to this problem and simulate non-preferential exploratory spatial behavior, 
with no artificial moving patterns or favoring of one segment of the environment over another, with speeds falling 
within experimental range. Such approach allows reproducing a natural flow of spiking data and estimating how long it 
takes to integrate it into a topological map. The statistical alternatives for the model are produced by randomizing 
place field maps over a fixed trajectory rather than by sampling over different trajectories, which is practically 
much more efficient.

It should be emphasized however, that these and all the subsequent simplifications should not be viewed as intrinsic 
limitations of the approach but only as approximations used for simplifying specific computations. The model would 
also work with more detailed information, e.g., using more precisely estimated spike times or behavioral parameters, 
physiologically recorded or generated via accurate network models, detailed synaptic transmission mechanisms, etc. 

\section{Overview of the results}
\label{section:results}

\begin{figure} 
	\includegraphics[scale=0.8]{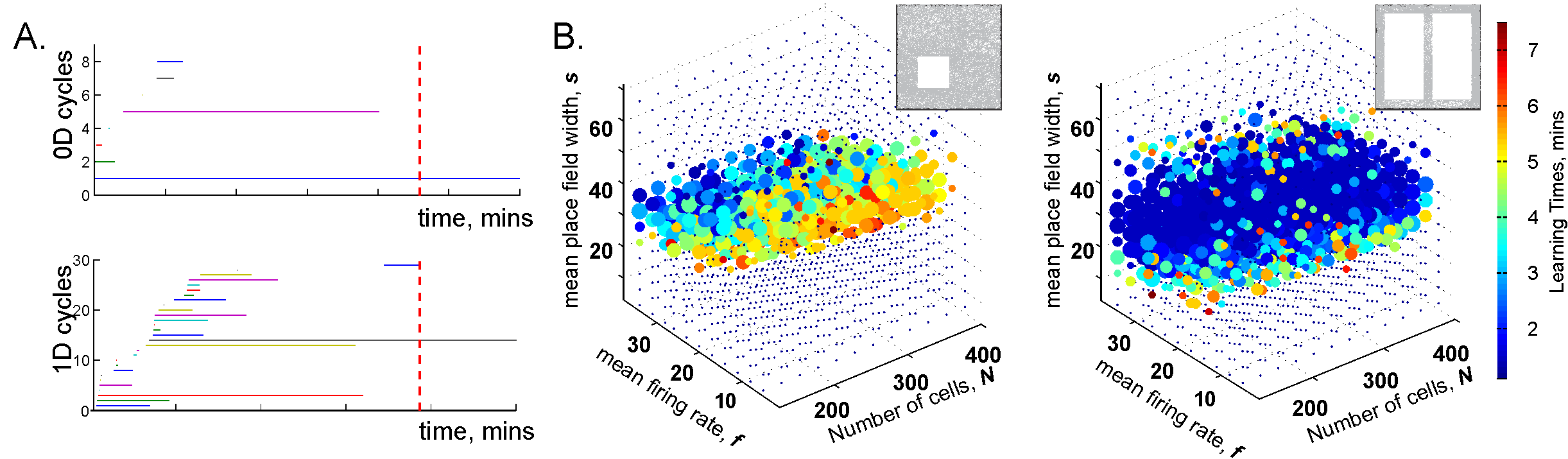}
	\caption{\textbf{Topological description of spatial learning}. \textbf{A}. As the rat begins to explore an 
		environment, the simplicial complex $\mathcal{T}(t)$ consists of $0D$ cycles that correspond to small groups 
		of cofiring cells that mark contractible spatial domains and $1D$ cycles that represent transient holes in
		$\mathcal{T}(t)$. As exploration continues, the place cell coactivity contributes more simplexes into 
		$\mathcal{T}(t)$, leading to disappearance of spurious cycles and leaving only a few persisting ones, which 
		express stable topological information \cite{Zomorodian2,Edelsbrunner,Ghrist2}. 
		\textbf{B}. Each point in the parameters space with coordinates $(s,f,N)$ represents a particular place cell 
		ensemble. The colors of the dots represent the mean learning time $T_{min}$ and their sizes represent the 
		success rate: small dots represent ensembles that only occasionally converge on the correct information, large 
		dots represent ensembles that converge most or all of the time. The ensembles that can produce a correct map 
		time occupy a particular domain of the parametric space---the \textit{Learning Region}, $\mathcal{L}$, where 
		learning is fastest and most accurate; near the boundary, map formation times become longer and the error rate 
		(failure to converge) increases. Outside of $\mathcal{L}$ learning fails. Importantly, the parameter values 
		that correspond to $\mathcal{L}$ happen to parallel experimentally derived values, which indicates a biological 
		relevance of the model. The smaller $\mathcal{L}$ on the left panel corresponds to a $2D$ arena with a hole, 
		about $1.5\times 1.5$ m in size, and larger $\mathcal{L}$ on the right corresponds to a topologically simple 
		quasi-linear environment (top right corner of each panel). The more complex the environment, the more tuned 
		the neural ensembles have to be to learn the space.}
	\label{Figure1}
\end{figure}

1. \textbf{The learning region}. For a particular set of values $(s, f, N)$, a trajectory traversing a place field 
map $M_{\mathcal{E}}$ produces a certain time-dependent coactivity complex $\mathcal{T}(t)$. One may thus inquire
whether, and for which ensembles, such a complex acquires the correct topological shape and how long this process 
may take. As it turns out, the coactivity complexes produced by generic place field maps can assume correct topological 
shapes, $b_k(\mathcal{T})=b_k(\mathcal{E})$, $k\geq 0$, in a biologically feasible period---if the spiking parameters 
fall into a specific domain in the parameter space that we refer to as the \textit{learning region}, $\mathcal{L}$ 
(Fig.\ref{Figure1}B). 
It is important to note that although the exact structure of $\mathcal{T}(t)$ depends profusely on the details of 
the map $M_{\mathcal{E}}$ \cite{SchemaM}, most large-scale characteristics of $\mathcal{T}(t)$, e.g., the overall
structure of its Betti numbers, are largely $M_{\mathcal{E}}$-independent. This leads to the model's first predictive 
outcome, namely to the observation that the mean spiking parameters $(s,f,N)$ may be used to identify a particular 
hippocampal `state' with a certain learning capacity \cite{PLoS}.

%%%%%%%%%%%%%%%%%%%%%%%%%%%%%%%%%%%
%%%%%%%%%%%%%%%%%%%%%%%%%%%%%%%%%%%

The second key observation is that the placement of the learning region in the parameter space matches the biological 
range of spiking characteristics derived from electrophysiological experiments. \textit{A priori}, this correspondence 
is not guaranteed: the region $\mathcal{L}$ that emerges from the `homological' computations could have appeared anywhere 
in the parameter space. However, the fact that the `operational' domain of the topological model appears to match the 
biological domain suggests that the topological approach captures some actual aspects of the neurophysiological 
computations taking place in the hippocampal network. In particular, it indicates that the physiological neurons can 
indeed encode a topological map of space in a biologically feasible time. On the other hand, boundedness of $\mathcal{L}$ 
also shows that spatial selectivity of firing does not, by itself, guarantee a reliable mapping of the environment, 
despite a widespread belief among neuroscientists to the contrary.

Third, the size and the shape of $\mathcal{L}$ reflect the scope of the biological variability that the hippocampus 
can afford in a given environment: the larger the learning region $\mathcal{L}$, the more stable the map (Fig.\ref{Figure1}B). 
Indeed, the model implies that the hippocampus can change its operating state inside $\mathcal{L}$ without compromising 
the integrity of the topological map: if one parameter begins to fall outside the learning region, then a successful 
spatial learning can still occur, provided that compensatory changes of other parameters can keep the neuronal ensemble 
inside $\mathcal{L}$. This observation allows reasoning about the effects of certain diseases (e.g., Alzheimer's 
\cite{Cacucci,Cohen}) or environmental toxins (e.g., ethanol \cite{White,Matthews}, cannabinoids \cite{Robbe}) that 
produce more diffuse place fields, lower place cell firing rates, smaller numbers of active cells and thus may disrupt 
spatial learning by shifting system's parameters beyond the perimeter of the learning region.

Fourth, the structure of the learning region may also vary with the geometry of the environment, the laboriousness 
of navigation: the greater the task's complexity, the narrower the range that can sustain learning---as suggested by 
experimental studies \cite{Nithianantharajah,Fenton1,Eckert}. Thus, despite the topological nature of the information 
processing, the place cells are not `agnostic' about the scale and the shape of the navigated space. In fact, it can 
be shown that maps of large spaces can be assembled from the maps of their parts, e.g., if a domain $\mathcal{E}$ is 
split into two subdomains $\mathcal{E}_1$ and $\mathcal{E}_2$ that meet but do not overlap, then one can compute the 
individual learning times $T_{\min}(\mathcal{E}_1)$ and $T_{\min}(\mathcal{E}_2)$ using only the spikes fired within 
each subdomain. The sum of these learning times is similar to total time spent by rat in the entire arena, 
$T_{\min}(\mathcal{E}) \approx T_{\min}(\mathcal{E}_1) + T_{\min}(\mathcal{E}_1)$, with statistically insignificant 
differences \cite{Arai}. 
Mathematically, this result may be viewed as an adaptation of the Mayer-Vietoris theorem that states that if a space 
$\mathcal{E}$ is split into pieces $\mathcal{E}_1$ and $\mathcal{E}_2$ that overlap over a domain with vanishing 
homologies, $H_q(\mathcal{E}_1\cap \mathcal{E}_2) = 0$, then the homologies of the whole space are given by the direct 
sum of the homologies of the components, $H_q(\mathcal{E}) = H_q(\mathcal{E}_1)\oplus H_q(\mathcal{E}_2)$ \cite{Hatcher}. 
In case of the coactivity complexes, simulations demonstrate that persistent loops that represent topological obstacles 
in two complementary domains combine into the set of the persistent loops that represent the whole space, providing a 
novel perspective on the learning process.

Note, that these outcomes of the model correspond well with our subjective learning experiences: the complexity of the 
task influences learning time; difficult tasks are accomplished at or just beyond the limits of our capacity; disease 
or intoxication can reveal limits in our spatial cognition that would normally be compensated for, and so forth.

2. \textbf{Coactivity window}. The results discussed above are based on topological analyses of spiking data produced 
by large populations of coactive place cells; but what defines neuronal coactivity in the first place?
At a phenomenological level, an instance of coactivity may be characterized by the length of the period allocated for
detecting the spikes fired by two or more cells. Experimental studies suggest that the `physiological' width $w$ of the 
coactivity window ranges between tens to hundreds of milliseconds, with the standard estimate $w \sim 200$ msec 
\cite{Ang,Maurer,Mizuseki,Huhn}. The topological model allows addressing this question theoretically: one can ask, e.g., 
what range of window sizes \textit{could} allow constructing topological maps and would these values match the biological 
range of coactivity periods? One can also inquire, given a particular width $w$, whether the dynamics of $\mathcal{T}(t)$ 
depends on a specific arrangement of the coactivity intervals along the time axis and how sensitive the results may be 
with respect to the windows' variations from one instance of coactivity to another. In biological terms: can the exact 
sequence and the variability of coactivity readouts affect the animal's learning capacity?

%%%%%%%%%%%%%%%%%%%%%%%%%%%%%%%%%%%%%%%
%\begin{figure} 
\begin{wrapfigure}{c}{0.5\textwidth}
	\includegraphics[scale=0.89]{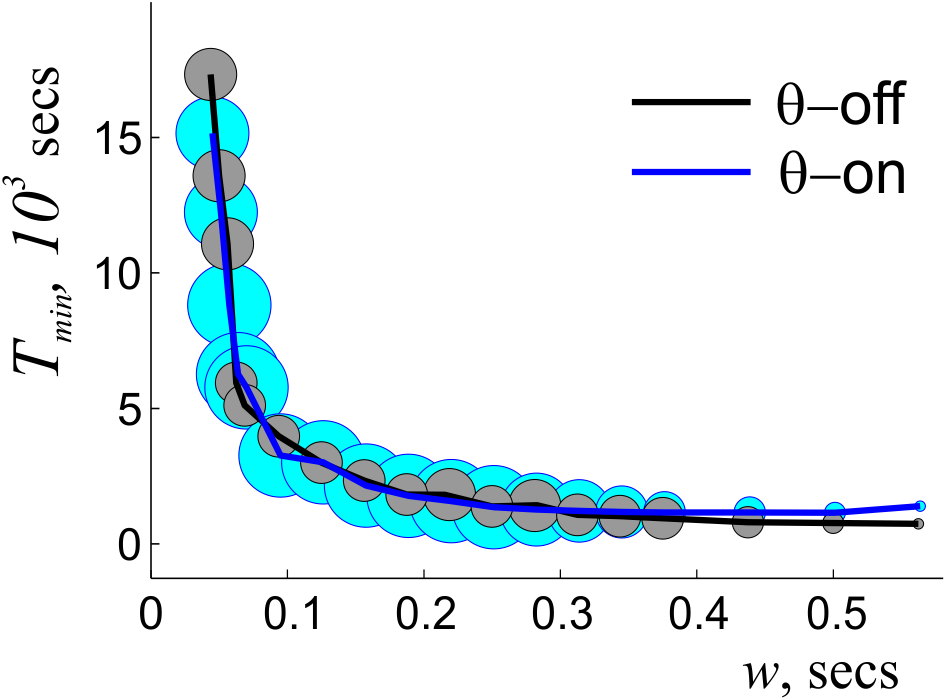}
	\caption{\label{Figure2} 
		{\footnotesize
			\textbf{Dependence of learning time on window width}, with and without $\theta$-modulation of spiking activity. 
			The radius of the circles indicates the percentage of times the coactivity complex assumes the correct topological 
			shape. In both cases, the coactivity complexes with the correct topological shape start to form at $w_o = 0.2$ 
			$\theta$-periods, when the learning time is long (hours) and sensitive to the variations of $w$, and fail at 
			$w\sim 4.5$ $\theta$-periods, when learning becomes unreliable. At $w_s\sim 1.5$ $\theta$-periods the dependence
			$T_{\min}(w)$ plateaus, marking the domain of stable $w$. Here $N = 350$, $f=28$ Hz, and $s = 23$ cm.
		}
	}
\end{wrapfigure}
%\end{figure} 
%%%%%%%%%%%%%%%%%%%%%%%%%%%%%%%%%%%%%%%

As it turns out, the answer to the latter two questions is negative: the statistics of place cell coactivity and hence 
the structure of the coactivity complex do not exhibit strong dependence on either the coactivity windows' random temporal 
shifts or on the window sizes' `jitter' (both for up to $50\%$ of the mean $w$). On the one hand, this justifies using 
a single parameter $w$ for studying the dependence of the coactivity complex' structure on the window width. On the other 
hand, it is clear that learning dynamics should depend on the systematic changes of $w$: if the coactivity window is too 
narrow, then the spike trains produced by the place cells will often `miss' one another, so that the map will either fail 
or take a long time to emerge. However, if $w$ is too wide, then the place cells with disconnected place fields will 
contribute spurious links that may compromise the map's structure. 

Simulations show that indeed, an accurate topological map emerges within a well-defined range of $w$s, $w_o \approx 25\leq w 
\leq w_c \approx 1250$ msec, beyond which the maps have vanishing convergence rates (i.e., maps rarely or never produces the 
correct Betti numbers). In-between, the learning time follows a power law dependence, $T_{\min}(w)\sim w^{-1.2}$, starting at 
high values ($T_{\min}(w_o) \approx 5$ hours) that rapidly decrease with growing $w$ (Fig.~\ref{Figure2}). 
However, the `operational' range of $w$s is even smaller since the biological dependence $T_{\min}(w)$ should also be 
\textit{stable}, i.e., it should not be hypersensitive to variations of $w$ or exhibit low convergence rates. In the model, 
such a range of $w$s lays approximately between $125$ and $250$ msec (Fig.~\ref{Figure2}), which matches the domain implicated 
in experimental studies. Thus, the model once again allows deriving the physiological values---in this case the widths of the 
coactivity windows---from purely theoretical considerations.

\begin{figure}%[!hbt] 
	\includegraphics[scale=0.8]{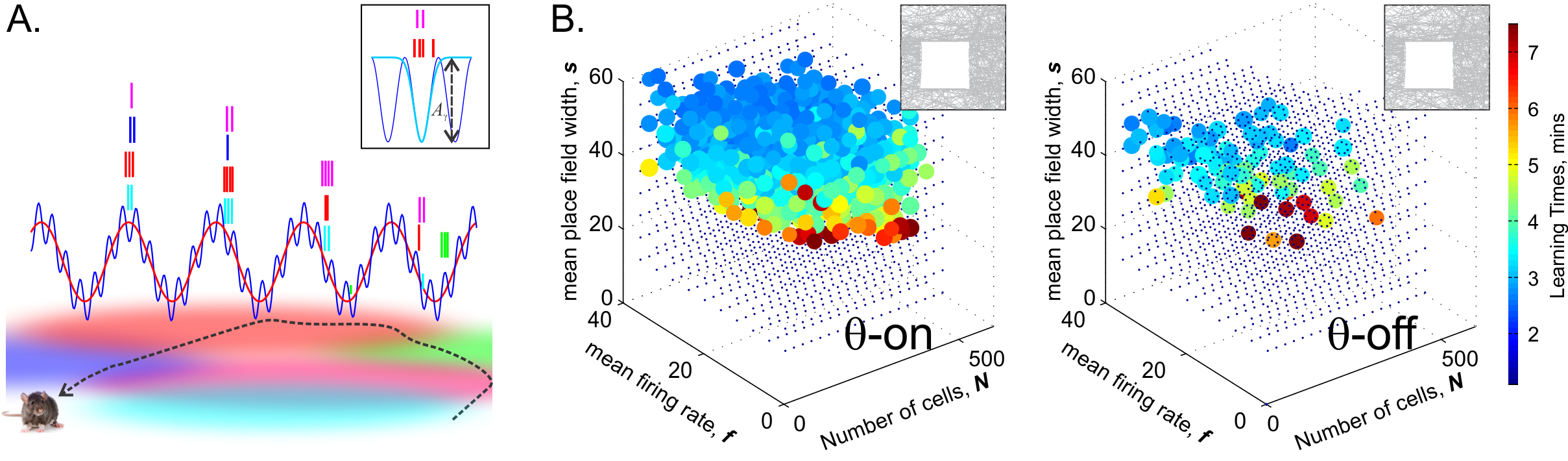}
	\caption{\textbf{Brain waves enhances learning}. \textbf{A}. $\theta$-precession and $\gamma$-synchronization 
		modulate place cell spiking activity: Spike times precess with the $\theta$-rhythm ($\approx 8$ Hz), shown
		as red wave: as the rat traverses a place field, the corresponding place cell discharges at a progressively 
		earlier phase in each new $\theta$-cycle. The preferred $\theta$-phases correspond to $\gamma$-cycles 
		($\approx 60$ Hz). The blue wave shows the net $\theta+\gamma$ amplitude. The spikes, shown by tickmarks 
		colored according to the place fields traversed by the animal's trajectory, cluster over the $\gamma$-troughs, 
		yielding dynamical cell assemblies. Boxed image: Spikes spread around a $\gamma$-trough of depth $A_{\gamma}$ 
		(dark blue curve) distribute similarly to stochastic particles in a $1D$ potential wells (light blue curve). 
		The spike time probabilities are modulated by a Boltzmann factor $e^{-A_{\gamma}(t)/\tau}$, where $\tau$ is 
		an `effective temperature.'
		\textbf{B}. Learning regions with $\theta$-precession (left) and without it (right). In the latter case, 
		the size and the density of $\mathcal{L}$ diminishes, indicating that $\theta$-oscillations enhance place 
		cells' ability of to encode spatial maps, affording them greater resilience in the face of the spiking rate 
		or population size changes. Computations are made for a $1\times 1$m environment own in the top left corner.}  
	\label{Figure3}
\end{figure}

3. \textbf{The brain waves}. The temporal organization of the spike trains is strongly influenced by the oscillating
extracellular electrical fields---the \textit{brain waves}, that control the temporal architecture of the spiking activity 
and the parcellation of the information flow in the brain \cite{Draguhn}. In particular, the $\theta$-wave (4-12 Hz) and the 
$\gamma$-waves (40-80 Hz), are known to modulate the place cells' activity at several timescales and 
affect spatial leaning \cite{BuzsakiTheta1,Hasselmo,Colgin1}. However, it remains unclear at what level, and through 
what mechanisms, do these waves exert their influence. Most theoretical analyses address the effect of $\theta$- and 
$\gamma$-rhythms on individual cells' spiking \cite{Jensen1,Lisman1,Hasselmo}. In contrast, the topological model allows 
addressing this question at the ensemble level, by tracing how the $\theta$- and the $\gamma$-modulation of spike trains 
changes the dynamics of the corresponding coactivity complexes, e.g., the speed of their convergence towards correct 
topological shape, the statistics of topological defects exhibited during this process and so forth.

i.	$\theta$-\textit{phase precession} is a key mechanism by which the $\theta$-wave controls place cell's spiking: as 
a rat progresses through a place field, the corresponding place cell spikes near a certain preferred $\theta$-phase that 
progressively diminishes for each new $\theta$-cycle \cite{BuzsakiTheta2,Huxter} (Fig.\ref{Figure3}A). As discussed in
\cite{Skaggs,Jensen3}, this phenomenon helps to recapitulate the temporal sequence of the rat's positions in space during 
each $\theta$-period and it is therefore widely believed to enhance learning \cite{BuzsakiTheta1,BuzsakiTheta2}.

Simulations show that indeed, $\theta$-precession significantly enlarges the learning region, making otherwise poorly 
performing ensembles much more capable of learning. Without $\theta$-precession, the learning region $\mathcal{L}$ is 
small and sparse, and vice versa, certain place cell ensembles that in absence of $\theta$ lay beyond the learning 
region, become functional with the addition of $\theta$-precession (Fig.\ref{Figure3}B). Moreover, $\theta$-precession 
increases the probability of convergence to the correct shape across all ensembles that can form accurate maps, which 
suggests that $\theta$-precession may not just correlate with, but actually enforce spatial learning \cite{Arai}. 

%%%%%%%%%%%%%%%%%%%%%%%%%%%%%%%%%%%
%%%%%%%%%%%%%%%%%%%%%%%%%%%%%%%%%%%

In terms of the coactivity complex' structure, $\theta$-enhancement of learning is manifested through shortened
durations of the spurious $1D$ cycles, %which control path connectivity of the neuronal representation of $\mathcal{E}$, 
while initially increasing their number. In other words, $\theta$-modulation suppresses spurious defects in the 
cognitive map at the price of creating more transient errors at the initial stages of the navigation. Curiously, 
simulations also show that learning times are relatively insensitive to the details of the $\theta$-wave structure: 
the presence of a spike-modulating $\theta$-rhythm by itself is more important than a specific wave shape \cite{Arai}. 

As for the interplay with the coactivity parameters, the stabilization of the $T_{\min}(w)$ dependence is achieved 
at approximately the same range of $w$s as without the $\theta$-precession, at $w \sim 1-2 \theta$-cycles \cite{Arai}
(Fig.~\ref{Figure2}). Such recurrent matches between the preferred coactivity timescale and the $\theta$-timescale 
suggest that the interplay between  neuronal spiking and the parameters of animal's behavior (e.g., speed) required 
for optimal processing of topological information may \textit{define} the temporal domain of neuronal synchronization 
in the rat's hippocampal network.

Thus, $\theta$-modulated coactivity complexes provide a self-consistent description of the hippocampal network's 
function at the $\theta$-timescale, predicting \textit{inter alia} an optimal integration window for reading out 
the information and the temporal domain of synchronization. 

ii. $\gamma$-\textit{modulation of spiking}. As $w$ shrinks beyond the range predicted for the independently 
$\theta$-precessing place cells ($w < w_o$), spatial learning fails. Interestingly, this happens precisely at the 
timescale where complementary mechanisms of spike synchronization, driven by the second key component of the hippocampal 
brain waves---the $\gamma$-oscillations---are taking over \cite{Colgin2,BuzsakiGamma}. This raises question about 
whether an additional $\gamma$-synchronization of spiking could improve the predicted properties of the cognitive 
map, i.e., produce topologically correct coactivity complexes \cite{Dragoi}.

Physiologically, $\gamma$-wave represents fast oscillations of the inhibitory postsynaptic potentials. As its 
amplitude $A_{\gamma}(t)$ drops at a certain location, the surrounding cells with high membrane potential spike 
\cite{Jia,Nikoli,Lisman3}. As a result, each $\gamma$-trough defines the preferred $\theta$-phase of several cells, 
i.e., marks an ignition of a particular place cell combination, represented by a coactivity simplex. Computationally, 
coupling spike times with the $\gamma$-wave can be achieved by modulating neuronal firing rates with a Boltzmann 
factor $e^{-A_{\gamma}(t)/\tau_i}$. The parameter $\tau_i$ can be interpreted as an effective `temperature' that 
controls the temporal spread of spikes around the $i^{th}$ $\gamma$-trough: for large mean $\tau = \langle\tau_i 
\rangle$, the spikes are `hot,' i.e., spread diffusely near the $\gamma$-troughs and for small $\tau$ they `freeze' 
at them. In particular, the case in which the spike trains are uncorrelated with the $\gamma$-troughs corresponds 
to the limiting case of an `infinitely hot' hippocampus ($\tau = \infty$, e.g., the pure $\theta$-modulated cells 
discussed above). Meanwhile, the `physiological' effective temperature that describes the characteristic huddling 
of spikes within a $\gamma$-period observed in the experiments \cite{Colgin1,Colgin2} is comparable to the mean 
$\gamma$-amplitude, $\tau\approx\overline{A_{\gamma}}$.

The net effect of the $\gamma$-modulation on the coactivity complexes is as follows: as the effective temperature 
drops and the temporal spread of the spikes near the $\gamma$-troughs shrinks, the coactivity complexes produce 
fewer, faster-contracting spurious loops. In particular, at the `physiological' effective temperatures, 
$\gamma$-synchronized cognitive map can robustly capture the topology of the environment by integrating place cell 
coactivity at the $\gamma$-timescale, i.e., yield \textit{finite} learning times at $w < w_o$s, which provides a 
direct demonstration of the importance of the $\gamma$-synchronization at the systemic level.  

This result sheds light on the well-known correlation between successful learning and retrieval with the increase 
of the $\gamma$-amplitude in raised attention states \cite{Vugt,Moretti,Lundqvist,Trimper}. In particular, it helps 
understanding why suppression of the $\gamma$-waves induced, e.g., by psychoactive drugs \cite{Whittington} such as 
cocaine \cite{McCracken,Dilgen}, or arising due to neurodegeneration or aging \cite{Vreugdenhil,Lu}, usually correlates 
with learning impairments---according to the model, all these phenomena suppress map formation---or retrieval---at 
the $\gamma$ timescale. On the constructive side, the model suggests a new characteristics of the $\gamma$-synchronized 
spiking activity---the effective $\gamma$-temperature of spiking---that may be studied empirically and explained via 
neuronal mechanisms.

4. \textbf{Ramifications of coactivity complexes}. The predictions derived from the topologico-physiological 
constructions discussed above are not universal. For example, а direct application of the model to the case of the 
bats navigating $3D$ caves \cite{Ulanovsky,Yartsev} often produces dysfunctional coactivity complexes, with hundreds 
of persistent spurious loops---even for the experimentally observed parameters of spiking activity \cite{Hoffman}. 
On the one hand, this failure can be explained by the relatively high speeds of the bat's movements (over $2$ m/sec), 
which helps producing spurious coactivities between place cells with non-overlapping place fields \cite{Hoffman}. 
On the other hand, it also suggests that the very idea that place cells operate by responding to certain spatial 
domains (currently dominating in the field) may be only a simplified interpretation of their spiking mechanism, 
suitable for low speeds and basic environments. The model points out that deriving topological maps from such 
`passive responses' may, at higher speeds, generate mismatches between the spatial pattern of the `prearranged' 
place fields and the temporal pattern of the corresponding place cells' coactivities that are processed by the brain. 
In other words, the model suggests that processing place cell coactivities requires \textit{editing} the raw pools 
of place cell spiking data---a surprising conclusion because it appeals to reasoning beyond the model's original 
setup. In effect, it suggests that the hippocampal network should be \textit{wired} to highlight some place cell 
coactivities and suppress others, even though no explicit references to the networks' structure were made in the 
original Alexandrov-\v{C}ech construction. 

\begin{figure}%[h] 
	\includegraphics[scale=0.8]{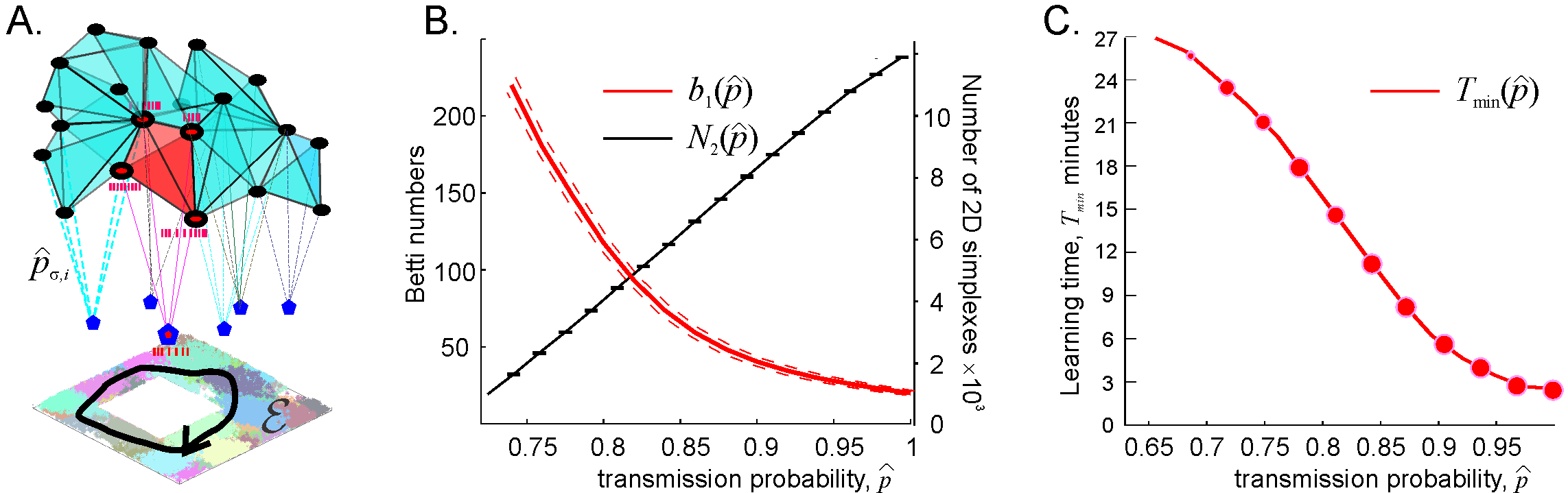}
	\caption{
		\textbf{Synaptic efficacies and cell assembly complexes}. 
		({\bf A}). Cell assemblies are functionally interconnected network of place cells (black dots) modeled as cliques 
		of the coactivity graph $\mathcal{G}$. Spikes from $k^{th}$ pair of coactive place cells in an assembly $\sigma$ 
		are transmitted to a readout neuron (blue pentagons) with probability $p_{\sigma,k}<1$ (an ignited cell assembly 
		is shown in red). The net structure of the cell assembly network is represented by the corresponding cell assembly 
		complex $\mathcal{T}_{CA}$, which captures the topology of underlying environment $\mathcal{E}$. 
		({\bf B}). The number of coactivity links shrinks with the diminishing spike transmission probability (black line) 
		at a power rate, whereas the number of spurious topological loops in $\mathcal{T}_{CA}$ proliferates exponentially.
		({\bf C}). As synapses weaken, the learning time $T_{\min}$ grows at a power rate. The size of the data points 
		represents the percentage of the outcomes with the correct Betti numbers ($b_{0}(\mathcal{T}_{CA}) = b_{1}(\mathcal{E}) 
		= 1$). Computations are performed using an ensemble of $N=400$ neurons with a mean firing rate of $f = 28$ Hz and mean 
		place field size $30$ cm.}  
	\label{Figure4}
\end{figure}

Curiously, this line of arguments addresses to a well-known neurophysiological phenomenon, namely the fact that place 
cells tend to form operative units known as \textit{cell assemblies}---functionally interconnected groups of neurons 
that drive their respective `readout' neurons in the downstream networks \cite{Harris1,Harris2,Syntax,Jackson,ONeill}. 
The spiking response of the latter actualizes connectivity relationships between the regions encoded by the individual 
place cells: if a specific instance of place cell coactivity does not elicit a response of a readout neuron, then the 
corresponding connectivity information does not contribute to the hippocampal map \cite{Syntax,SchemaS}. 
Thus, a cell assembly network of a specific architecture controls processing of the information supplied by the place 
cell spiking activity and the overall connectivity structure of the cognitive maps (Fig.~\ref{Figure4}A).

\textit{Clique coactivity complexes}. A simple model a place cell assembly network can be built by constructing a 
coactivity graph $\mathcal{G}$, whose vertexes $v_i$ correspond to place cells $c_i$ and the links, $\varsigma_{ij} = 
[v_i, v_j]$ represent the connections (functional or physiological) between pairs of coactive cells \cite{Muller,Burgess}. 
The place cell assemblies then correspond to fully interconnected subgraphs of $\mathcal{G}$, i.e., to its maximal cliques 
$\varsigma=[c_0,c_1,...,c_n]$. As a combinatorial objects, cliques are identical to the simplexes span by the same sets 
of vertexes; hence the collection of $\mathcal{G}$-cliques produces a complex \cite{Jonsson} that may serve as a 
schematic representation of either the cell assembly network or the cognitive map encoded by it \cite{CAs}. 

Simulations show that such complexes, denoted below as $\mathcal{T}_{\varsigma}$, are structurally very similar to 
the original coactivity complexes derived from the higher-order place cell coactivities, which we will denote as 
$\mathcal{T}_{\sigma}$. 
However functionally, $\mathcal{T}_{\varsigma}$s often perform much better than $\mathcal{T}_{\sigma}$s, e.g., they 
exhibit a much smaller number of shorter-living spurious loops, more robust learning times, etc. \cite{Basso,Hoffman,CAs}. 
The explanation for this effect is simple: the lowest order, \textit{pairwise} place cell coactivities are captured easier 
and more reliably than the higher-order coactivity events \cite{Katz,Brette}. An additional advantage is offered by a 
structural flexibility of the clique coactivity complexes, since it is possible to \textit{assemble} its individual 
cliques $\varsigma\in \mathcal{T}_{\varsigma}$ by \textit{accumulating} low order coactivities over time, rather than by
\textit{detecting} higher-order coactivity events. For example, in order to identify a third-order coactivity clique, 
$\varsigma = [c_i, c_j, c_k]$, one can first detect the coactive pair $[c_i, c_j]$, then the pair $[c_j, c_k]$ and then 
$[c_i, c_k]$, over an extended integration period $\varpi$, whereas in order to produce a coactivity simplex $\sigma = 
[c_i, c_j, c_k]$, all three cells must become active within a single coactivity window $w$. 

From the physiological perspective, the clique construction can be used to model a wide scope of physiological phenomena, 
e.g., for testing whether the readout neurons may operate as `coincidence detectors' that respond to nearly simultaneous 
activity of the presynaptic cells (for short integration windows $\varpi\sim w$ \cite{Katz,Brette}) or as `integrators' 
of the spiking inputs (for $\varpi\gg w$ \cite{Konig,Ratte,Magee1,London,Spruston}), along with the intermediate and/or 
mixed cases. The original approach based on the Alexandrov-\v{C}ech's construction corroborates with the first scenario: 
indeed, the nerve complex $\mathcal{N}$ is derived from the spatial overlaps between the regions, which mark the domains 
of nearly simultaneous place cell coactivity. The architecture of the clique coactivity complex suggests an alternative 
approach that significantly broadens the models' capacity to represent physiological phenomena.

%%%%%%%%%%%%%%%%%%%%%%%%%%%%%%%%%%%
%%%%%%%%%%%%%%%%%%%%%%%%%%%%%%%%%%%

Simulations show that, in fact, the connections within most cliques of $\mathcal{G}$ activate nearly simultaneously, i.e.,
most simplexes of $\mathcal{T}_{\sigma}$ are also present in $\mathcal{T}_{\varsigma}$. 
Nevertheless, a small population of cliques are never observed as simultaneous coactivity events and require assembling 
over extended periods \cite{Hoffman}. As a result, clique coactivity complexes $\mathcal{T}_{\varsigma}$ are typically 
larger and produce much fewer spurious topological loops that rapidly disappear with learning. In particular, such 
complexes produce correct topological maps of $3D$ spaces for the experimentally observed parameters of the spiking 
activity \cite{Hoffman}, suggesting that the readout neurons in bats' (para)hippocampal areas should function as 
integrators of synaptic inputs (with estimated spike integration period of about $4$ minutes), rather than detectors 
of place cells' coactivity---a prediction that may potentially be verified experimentally.

Another curious difference between the rats' and the bats' cognitive map construction mechanism is that less than 4\% of 
the bat's place cells exhibit significant $\theta$-modulated firing \cite{Yartsev}, which implies that $\theta$-precession 
in these animals may not play the same role as in rats. 
Indeed, simulating bat's movements with and without $\theta$-precession reveals that in the $\theta$-off case, the ensembles 
of place cells acquire correct maps faster than in the $\theta$-on cases, producing fewer topological loops both in the 
simplicial and in the clique coactivity complexes \cite{Hoffman}. To explain these results, one can consider the effect 
of $\theta$-precession from two perspectives: on the one hand, it synchronizes place cells and hence increases their 
coactivity rate, which may help learning \cite{BuzsakiTheta1,Lee,Jezek,Geisler,HarrisTheta}. 
On the other hand, it can be viewed as a constraint that reduces the probability of the cells' coactivity and hence 
decimates the pool of coactivity events. In relatively slow moving rats, when the coactivity events are reliably captured, 
the first effect dominates, contributing a steady influx of grouped spikes to downstream neurons. In rapidly moving bats 
however, when the network struggles to capture the coactivity events, the constraint imposed by phase precession acts as 
more of an impediment and slows down spatial learning process. Thus, the model once again allows using theoretical reasoning 
to produce a functional insight into the neurophysiological properties of the network from the parameters of task that it 
solves.

5.	\textbf{Cell assembly complex}. Question arises, whether the coactivity complexes may be implemented, in some 
capacity, in physiological networks or vice versa, whether it is possible to construct complexes that capture the 
organization of the cell assembly networks. Simulations show that the original set of coactive place cell combinations 
is very large: the numbers of $d$-dimensional simplexes in $\mathcal{T}_{\varsigma}$, $N_d$, scale proportionally to the 
binomial coefficients $C^{d+1}_N$. More specifically, it can be shown that the ratios $\eta_d = N_d / C^{d+1}_N$ depend 
primarily on the mean place field sizes and the firing rates and not on the number of cells within the ensemble, $N$
\cite{Arai}. In contrast, the number of cells that may potentially serve as readout neurons is similar to the number 
of place cells $N$, which implies that only a small fraction of coactive place cell groups can form assemblies 
\cite{Syntax,Shepherd}. 
This raises the question: is it possible to identify an `operational' set of place cell combinations---putative cell
assemblies---using simple selection rules?

In model's terms, the task of identifying a subpopulation of coactive place cell combinations corresponds to selecting 
a ``cell assembly subcomplex'' $\mathcal{T}_{CA}$ of $\mathcal{T}_{\varsigma}$, according to some biologically motivated 
criteria. First, the total number of the maximal simplexes in $\mathcal{T}_{CA}$ should be comparable to the number of 
its vertexes, $N_{\max}(\mathcal{T}_{CA})\approx N(\mathcal{T}_{CA}$), but the latter should not differ significantly 
from the original number of place cells, $N(\mathcal{T}_{CA}) \approx N(\mathcal{T}_{\varsigma})$. 
Second, only a few cell assemblies (selected cliques) should be active at a given location, to avoid redundancy of the 
place cell code. Conversely, the periods during which all place cell assemblies are inactive should be short, so that 
the rat's movements should not go unnoticed by the hippocampal network. Third, the larger is the number of cells shared 
by consecutively igniting cell assemblies (i.e., by the adjacent simplexes in a simplicial path), the more contiguous is 
the representation of the rat's moves. Hence the contiguity between the simplexes in $\mathcal{T}_{CA}$ should not 
decrease compared to $\mathcal{T}_{\varsigma}$. Lastly, $\mathcal{T}_{CA}$ should correctly capture the topological 
shape of the environment \cite{CAs}.

As it turns out, it is possible to carry out the required construction by selecting the most prominent combinations 
of coactive place cells---the ones that appear most frequently. This selection principle is motivated by the Hebbian 
``fire together wire together" neuronal plasticity mechanisms: frequently appearing combinations have a higher chance of 
being wired into the network \cite{Neves}. Specifically, one can construct the desired clique complexes by identifying 
the connections the coactivity graphs $\mathcal{G}$($\xi$) that activate at a rate exceeding a certain threshold $\xi$. 
Alternatively, one can select, for every cell $c_i$, its $n_0$ neighbor-cells that are most frequently coactive with 
$c_i$, which yields another family of coactivity graphs, $\mathcal{G}(n_0)$. Computations show that the first family, 
$\mathcal{G}(\xi)$, exhibits certain random graph properties while the second family, $\mathcal{G}(n_0)$, demonstrates 
scale-free properties \cite{Barabasi,Albert}, characteristic of the hippocampal network \cite{Li,Bonifazi}. However,
both families of ``restricted'' coactivity graphs allow constructing operational cognitive map models, for a viable 
set of $\xi$s and $n_0$s.

As expected, the size and the dimensionality of the corresponding clique complexes, $\mathcal{T}_{\varsigma}(\xi)$ and 
$\mathcal{T}_{\varsigma}(n_0)$, decrease with the growing threshold $\xi$ or diminishing $n_0$. In addition, their 
maximal simplexes become more contiguous and their number, $N_{\max}$, remains close to the number of cells. Lastly, 
the topological behavior of both $\mathcal{T}_{\varsigma}(\xi)$ and $\mathcal{T}_{\varsigma}(n_0)$ is also regular: 
with minor rectification algorithms that do not produce significant changes of the complex's structure or alter the 
appearance rate of simplexes, correct topological shapes can be attained as fast and as reliably as with the entire 
set of the place cell coactivities, without compromising the place cell coverage of the environment or fragmenting the
map \cite{CAs}. Thus, the generic biological requirements listed above are met and we may conclude that the selected 
`critical mass' of coactive place cell combinations can produce viable cell assembly complexes $\mathcal{T}_{CA}(\xi)$ 
and $\mathcal{T}_{CA}(n_0)$ \cite{CAs}.

6.	\textbf{Synaptic parameters}. The physiologically implementable cell assembly complexes $\mathcal{T}_{CA}$ set the 
stage for further developments of the topological model. For example, the simplexes of $\mathcal{T}_{CA}$ can be ``rigged'' 
with parameters describing transferring, detecting and interpreting neuronal (co)activity in the corresponding cell 
assemblies, allowing us to account for the effects of the hippocampal network's synaptic architecture and providing 
a basic description of the synaptic computations in the cell assemblies.

In a phenomenological approach, synaptic connections can be characterized simply by the probabilities of transmitting 
spikes from a place cell to a readout neurons' membranes and by the probabilities that the latter will spike upon 
collecting their inputs. If the cell assemblies are modeled as cliques of the coactivity graph, the key role is played 
by the probability of transmitting the coactivity from the pairs of coactive place cells to the corresponding readout 
neurons' and response probabilities. 
In principle, these probabilities can be evaluated using detailed neuronal and synaptic models; however, in a simpler 
phenomenological approach, they may be regarded as random variables drawn from stationary, unimodal distributions with 
the modes $p_{\ast}$ (transition) and $q_{\ast}$ (response) and variances $\Delta_p$ and $\Delta_q$. The stationarity 
here implies that we disregard synaptic plasticity processes \cite{BuzLog,Barbour,Brunel}.

Under such assumptions, it is possible to study how the large scale, systemic characteristics of the spatial memory 
map depend on the synaptic strengths, at what point spatial learning may fail, and so forth. It can be shown, e.g., 
that if the characteristic coactivity transmission probability is high ($0.9 \leq p_{\ast} \leq 1$), then its small 
variations do not produce strong effects on the spatial map. On the other hand, as $p_{\ast}$ decreases further, the 
changes accumulate and, as $p_{\ast}$ approaches a certain critical value $p_{crit}$, learning times diverge at a power 
rate, $$T_{\min}\propto(p_{\ast}-p_{crit})^{-\kappa},$$ with $\kappa$ ranging typically between $0.1$ and $0.5$. The 
effects produced by the diminishing probability of the postsynaptic neurons' responses, $q_{\ast}$, are qualitatively 
similar but weaker than the effects of lowering the spike transmission probability $p_{\ast}$ \cite{Efficacies}. 

These results suggest explanations for numerous observations of correlative links between weakening memory capacity 
and deterioration of synapses, broadly discussed in neuroscience literature \cite{Selkoe,Toth}. According to model, 
weakening synapses reduce the size the coactivity complex and degrade its topological structure. For example, 
simulations demonstrate the number of connections in the coactivity graph $\mathcal{G}$ near $p_{crit}$ drops as 
$N_2\propto(p_{\ast}-p_{crit})^{\delta}$, $\delta \sim 1$, whereas the number of longer-lasting $1D$ spurious loops in 
the corresponding coactivity complex grows exponentially, $\log(b_1)\propto (p_{crit}-p_{\ast})$ (Fig.~\ref{Figure4}B), 
suggesting a phase transition from a regular to an irregular state \cite{Vaccarino,Ambjorn}. In physiological terms, 
this implies that synaptic depletion reduces the number of detectable coactivities, while generating defects in the 
cognitive map, which results in a rapid increase of the learning time. 

Moreover, weakening synapses reduce the learning region down to its compete disappearance at $p_{\ast} = p_{crit}$, 
which suggests that spatial learning may fail not only because the parameters of neuronal firing are pushed beyond 
a certain fixed `working range,' but also because that range itself may shrink or cease to exist. In particular, the 
fact that the learning region disappears if the transmission probability drops below the critical value implies that 
deterioration of memory capacity produced by the synaptic failure cannot be compensated by increasing the place field's 
firing rates or by recruiting a larger population of active neurons---for more details see \cite{Efficacies}. 

7. \textbf{Memory spaces}. In the above discussion, the coactivity complexes were used to describe topological structure
of the hippocampal spatial memory frameworks---cognitive maps \cite{Moser,Schmidt}. However, it is well known that 
hippocampus encodes not only spatial but also generic, nonspatial memories \cite{Wood,Ginther,Wixted}, embedding them 
into broader contexts, placing them in sequence of preceding and succeeding events \cite{Agster,Fortin1}. 
In \cite{EichMem} it was suggested that the resulting integrated memory structure may be viewed as a \textit{memory space}
$\mathcal{M}$ that subjects can `mentally explore' or even `mentally navigate' \cite{Hopfield}. In other words, it was 
suggested that individual memory episodes and the spatiotemporal relationships between them may be viewed as `locations' 
or `domains' that may overlap, contain one another or be otherwise related in a `spatial' manner \cite{SchemaM}. In 
particular, the standard spatial inferences that enable spatial cognition and behavior are viewed as particular examples 
of the memory space navigations \cite{Hopfield,Issa,Novikov}.

From a physiological perspective, the fact that а memory space associated with a given environment $\mathcal{E}$ is 
encoded by the same place cell population that produces a cognitive map of $\mathcal{E}$, suggests that the corresponding 
coactivity complex $\mathcal{T}_{CA}$ may be used to represent both structures. To gain an insight into this representation, 
notice that $\mathcal{T}_{CA}$ defines a finite topological space $\mathcal{A}(\mathcal{T}_{CA})$, endowed with Alexandrov 
topology: the locations in $\mathcal{A}(\mathcal{T}_{CA})$ correspond to the coactivity simplexes and the topological 
neighborhoods of a given location represented by a simplex $\varsigma$ are formed by the locations whose simplexes 
include $\varsigma$ \cite{Alexandroff,SchemaM}. Since the simplexes of $\mathcal{T}_{CA}$ represent combinations of 
coactive place cells, which, in turn, are believed to represent memory elements, one may view the resulting `topological 
space of coactivities' $\mathcal{A}(\mathcal{T}_{CA})$ as a representation of the topological memory space encoded by 
the corresponding cell assembly network, $\mathcal{M}(\mathcal{T}_{CA})$. There are three immediate implication of this 
construction.

i. The dynamics of the large-scale topological structure of memory space can be translated directly from the 
algebro-topological studies of the corresponding coactivity complexes, since the (singular) homologies of 
$\mathcal{M}(\mathcal{T}_{CA})$ are identical to the (simplicial) homologies of the coactivity complex $\mathcal{T}_{CA}$ 
\cite{CAs,McCord,Stong}). This implies, e.g., that a memory space that contains a topological map of a given environment 
emerges over the same learning period $T_{\min}$ and within the same scope of spiking parameters $\mathcal{L}$ as the 
cognitive map, that it is similarly affected by the brain waves, by the deteriorating synapses and so forth.

ii. It can be shown that neuronal activity representing a trajectory $\gamma$ traced by the animal in physical space maps 
continuously into path $\wp$ navigated in the Alexandrov topology of the memory space $\mathcal{M}(\mathcal{T}_{CA})$. This
provides a theoretical base for the intuition of `mental exploration', by allowing to interpret the succession of the place 
cell activities as a representation of a continuous succession of memory episodes \cite{Samsonovich,Buzsaki2,Issa,Dabaghian1}.

iii. In neuroscience literature it is recognized that ``\textit{space is constructed in the brain rather than perceived, and 
the hippocampus is central to this construction}," and yet its meaning remains unclear: ``\textit{how can spaceless data enter 
the hippocampal system and spatial cognitive maps come out}" \cite{Nadel,OKeefe}. The topological model may shed light on these 
problems, because it allows interpreting spatiality \textit{intrinsically}, as a certain \textit{relational} structure defined 
on spiking activity \cite{Cohn,Roeper,Vickers}, thus providing an ontological foundation for the emergent spatiality of the
cognitive map, discussed in the Introduction.

\section{Discussion}
\label{section:discussion}

Extensive studies are dedicated to establishing correlations between parameters of neuronal activity and the 
characteristics of cognitive phenomena that emerge from this activity \cite{Postle}. In particular, decades of 
research were dedicated to relating spiking parameters of the hippocampal place cells, the synaptic architecture 
of the hippocampal network to the animal's ability to find paths and shortcuts, to remember how to evade obstacles, 
to retain and retrieve memories etc. The approach discussed above aims at filling the ``semantic gap" between these 
two scales of information processing within a unified framework, based on the physiological conjecture about a 
topological nature of the hippocampal memory structures \cite{eLife,SchemaM,SchemaS}. 

The connection with the realm of simplicial topology used in this framework is made based on an observation that 
neuronal computations may be formally described as operations over spikes combinations---which ones are produced 
over a given period, which ones are detected or transformed into specific outputs, etc. Viewing each particular 
collection of spikes as an abstract simplex allows representing larger volumes of spiking data as abstract 
simplicial complexes whose topological properties describe the net qualitative information emerging at the ensemble 
level. With this approach, the simplicial complex' dynamics may be used as a metaphor for the learning and other 
cognitive processes, which permits not only phenomenological descriptions at different spatiotemporal scales but 
also possesses explanatory power, i.e., allows embedding empirical data into qualitative and quantitative schemas 
for reasoning about physiological phenomena.


\newpage
\begin{thebibliography}{99}

\bibitem{OKeefe} O'Keefe J. \& Nadel, L. \textit{The hippocampus as a cognitive map.} Oxford University Press (1978).

\bibitem{Moser} Moser, E., Kropff, E. \& Moser, M-B. Place Cells, Grid Cells, and the Brain's Spatial Representation System. \textit{Annu Rev Neurosci} \textbf{31}: 69-89 (2008).

\bibitem{Schmidt} Schmidt, B. \& Redish, A. Neuroscience: Navigation with a cognitive map. \textit{Nature} \textbf{497}: 42-43 (2013).
\bibitem{Agarwal} Agarwal, G., Stevenson, I., Ber{\'e}nyi, A., Mizuseki, K., and Buzs{\'a}ki, G. \& Sommer, F. Spatially Distributed Local Fields in the Hippocampus Encode Rat Position. \textit{Science} \textbf{344}: 626-630 (2014).

\bibitem{Gothard} Gothard, K., Skaggs, W. \& McNaughton, B. Dynamics of mismatch correction in the hippocampal ensemble code for space: interaction between path integration and environmental cues, \textit{J Neurosci.} \textbf{16}: 8027-8040 (1996).
\bibitem{Leutgeb} Leutgeb, J., Leutgeb, S., Treves, A., Meyer, R., Barnes, C., McNaughton, B., Moser, M-B. \& Moser, E. Progressive transformation of hippocampal neuronal representations in ``morphed'' environments, \textit{Neuron}, \textbf{48}: 345-358 (2005).
\bibitem{Wills} Wills, T., Lever, C., Cacucci, F., Burgess, N. \& O'Keefe, J. Attractor dynamics in the hippocampal representation of the local environment, \textit{Science}, \textbf{308}: 873-876 (2005).
\bibitem{Touretzky} Touretzky, D., Weisman, W., Fuhs, M., Skaggs, W., Fenton, A. \& Muller, R. Deforming the hippocampal map. \textit{Hippocampus}, \textbf{15}: 41-55 (2005).
\bibitem{eLife} Dabaghian, Y., Brandt, V. \& Frank, L. Reconceiving the hippocampal map as a topological template, \textit{eLife} \textbf{10}.7554/eLife.03476: 1-17 (2014).
\bibitem{Alvernhe} Alvernhe, A., Sargolini, F. \& Poucet, B. Rats build and update topological representations through exploration, \textit{Anim. Cogn.}, \textbf{15}: 359-368 (2012).
\bibitem{Poucet} Poucet, B. \& Herrmann, T. Exploratory patterns of rats on a complex maze provide evidence for topological coding. \textit{Behav Processes}, \textbf{53}: 155-162 (2001).
\bibitem{Wu} Wu, X. \& Foster, D. Hippocampal replay captures the unique topological structure of a novel environment. \textit{J. Neurosci.}, \textbf{34}: 6459-6469 (2014).

\bibitem{Wilson} Wilson, M.A. \& McNaughton, B.L. Dynamics of the hippocampal ensemble code for space. \textit{Science} \textbf{261}: 1055-1058 (1993).
\bibitem{Pouget} Pouget, A., Dayan, P. \& Zemel, R. Information processing with population codes. \textit{Nature Rev. Neurosci.} \textbf{1}: 125-132 (2000).
\bibitem{Postle} B. Postle, Working Memory as an Emergent Property of the Mind and Brain, \textit{Neuroscience} \textbf{139}: 23-38 (2006).%(1)

\bibitem{Alex} Alexandroff, P. Untersuchungen \"{u}ber Gestalt und Lage abgeschlossener Mengen beliebiger Dimension. \textit{Annals of Mathematics}, \textbf{30}: 101-187 (1928).
\bibitem{Cech} \v{C}ech, E. Th\'eorie g\'en\'erale de l'homologie dans un espace quelconque. \textit{Fundamenta mathematicae}, \textbf{19}: 149-183 (1932).
\bibitem{Hatcher} Hatcher, A. \textit{Algebraic topology. (Cambridge University Press 2002)}.

\bibitem{Ghrist1} De Silva, V. \& Ghrist, R.\textit{Coverage in sensor networks via persistent homology}. Algebraic \& Geometric Topology \textbf{7}: 339-358 (2007).
\bibitem{Curto} Curto, C. \& Itskov, V. Cell groups reveal structure of stimulus space, \textit{PLoS Comput. Biol.}, \textbf{4}: e1000205 (2008).

\bibitem{PLoS} Dabaghian, Y., M\'emoli, F., Frank, L. \& Carlsson, G. A Topological Paradigm for Hippocampal Spatial Map Formation Using Persistent Homology, \textit{PLoS Comput. Biol.}, \textbf{8}: e1002581 (2012).
\bibitem{Arai} Arai, M., Brandt, V. \& Dabaghian, Y. The effects of theta precession on spatial learning and simplicial complex dynamics in a topological model of the hippocampal spatial map, \textit{PLoS Comput. Biol.}, \textit{10}: e1003651 (2014).
\bibitem{Hoffman} Hoffman, K., Babichev, A. \& Dabaghian, Y. A model of topological mapping of space in bat hippocampus, \textit{Hippocampus}, \textbf{26}: 1345-1353 (2016).
\bibitem{Basso} Basso, E., Arai, M. \& Dabaghian, Y. The effects of $\gamma$-synchronization on spatial learning in a topological model of the hippocampal spatial map, \textit{PloS Comput. Biol.} \textbf{12}: 9 (2016). 
\bibitem{CAs} Babichev, A., M\'emoli, F., Ji, D. \& Dabaghian, Y. A topological model of the hippocampal cell assembly network, \textit{Frontiers in Comput. Neurosci.}, \textbf{10}: 50 (2016).
\bibitem{Efficacies} Y. Dabaghian (2018), Through synapses to spatial memory maps: a topological model, \textit{Sci. Rep}. \textbf{9}: 572 (2019).
\bibitem{SchemaM} Babichev,  A. \& Dabaghian, Y. Topological schemas of memory spaces, \textit{Frontiers Comput. Neurosci.} \textbf{12}: 27 (2018).
\bibitem{SchemaS} Babichev, A., Cheng, S. \& Dabaghian, Y. Topological schemas of cognitive maps and spatial learning. \textit{Front. Comput. Neurosci.}, \textbf{10}: 18 (2016). 
\bibitem{Novikov} Dabaghian, Y. Maintaining Consistency of Spatial Information in the Hippocampal Network: A Combinatorial Geometry Model, \textit{Neural Computation}, \textbf{28}: 1051-1071 (2016).

\bibitem{Brown1} Brown, E., Frank, L., Tang, D., Quirk, M. \& Wilson, M. A statistical paradigm for neural spike train decoding applied to position prediction from ensemble firing patterns of rat hippocampal place cells. \textit{J. Neurosci.}, \textbf{18}: 7411-7425 (1998).
\bibitem{Jensen1} Jensen, O. \& Lisman, J.E. Position reconstruction from an ensemble of hippocampal place cells: contribution of theta phase coding. \textit{J. Neurophysiol.} \textbf{83}: 2602-2609 (2000).
%\bibitem{Barbieri1} Barbieri, R., Wilson, M.A., Frank, L.M. \& Brown, E.N. An analysis of hippocampal spatio-temporal representations using a Bayesian algorithm for neural spike train decoding. \textit{IEEE transactions on neural systems and rehabilitation engineering} \textbf{13}: 131-136 (2005).
\bibitem{Guger} Guger, C., Gener, T., Pennartz, C., Brotons-Mas, J., Edlinger, G., Berm\'udez, I., Badia, S., Verschure, P., Schaffelhofer, S. \& Sanchez-Vives MV. Real-time position reconstruction with hippocampal place cells. \textit{Front. Neurosci.}, \textbf{5}: 85 (2011).

\bibitem{Ghrist2} Ghrist, R. Barcodes: The persistent topology of data, \textit{Bull. Amer. Math. Soc.}, \textbf{45}: 61-75 (2008).
\bibitem{Zomorodian2} Zomorodian, A., \& Carlsson, G. Computing persistent homology. \textit{Discrete Comput Geom} \textbf{33}: 249--274 (2005).
\bibitem{Edelsbrunner} Edelsbrunner, H. \& Harer, J. \textit{Computational topology: an introduction.} Amer. Math. Soc. (2010).

\bibitem{Barbieri2} Barbieri, R., Frank, L., Nguyen, D., Quirk, M., Solo, V., Wilson, M. \& Brown E. Dynamic analyses of information encoding in neural ensembles, \textit{Neural Comput.} \textbf{16}: 277-307 (2004).

\bibitem{FentonVar} Fenton, A. A. \& R. U. Muller. Place cell discharge is extremely variable during individual passes of the rat through the firing field. \textit{Proc. Natl. Acad. Sci.} \textbf{95}: 3182-3187 (1998). %(6)

\bibitem{BuzLog} Buzs\'aki, G. \& Mizuseki, K. The log-dynamic brain: how skewed distributions affect network operations. \textit{Nat. Rev. Neurosci.} \textbf{15}: 264-278. (2014). %(4)
\bibitem{Barbour} Barbour, B., Brunel, N., Hakim, V. \& Nadal, J.-P. What can we learn from synaptic weight distributions? \textit{Trends in neurosciences} \textbf{30}: 622-629 (2007).
\bibitem{Brunel} Brunel, N., Hakim, V., Isope, P., Nadal, J.-P. \& Barbour, B. Optimal Information Storage and the Distribution of Synaptic Weights: Perceptron versus Purkinje Cell. \textit{Neuron} \textbf{43}: 745-757 (2004).

\bibitem{Cacucci} Cacucci, F., Yi, M., Wills, T.J., Chapman, P. \& O'Keefe, J. Place cell firing correlates with memory deficits and amyloid plaque burden in Tg2576 Alzheimer mouse model. \textit{Proc. Natl. Acad. Sci.} \textbf{105}: 7863-7868 (2008). 
\bibitem{Cohen} Cohen, R., Rezai-Zadeh, K., Weitz, T., Rentsendorj, A., Gate, D., Spivak, I., Bholat, Y., Vasilevko, V., Glabe, C., Breunig, J., Rakic, P., Davtyan, H., Agadjanyan, M., Kepe, V., Barrio, J., Bannykh, S., Szekely, C., Pechnick, R. \& Town, T. A Transgenic Alzheimer Rat with Plaques, Tau Pathology, Behavioral Impairment, Oligomeric A$\beta$, and Frank Neuronal Loss. \textit{J. Neurosci.} \textbf{33}: 6245-6256 (2013).

\bibitem{White} White, A. \& Best, P. Effects of ethanol on hippocampal place-cell and interneuron activity. \textit{Brain Res.} \textbf{876}: 154-165 (2000).
\bibitem{Matthews} Matthews, D., Simson, P. \& Best, P. Ethanol alters spatial processing of hippocampal place cells: a mechanism for impaired navigation when intoxicated. \textit{Alcohol Clin. Exp. Res.} \textbf{20}: 404-407 (1996).
\bibitem{Robbe} Robbe, D. \& Buzs\'aki G. Alteration of theta timescale dynamics of hippocampal place cells by a cannabinoid is associated with memory impairment. \textit{J. Neurosci.} \textbf{29}: 12597-12605 (2009).

\bibitem{Nithianantharajah} Nithianantharajah, J. \& Hannan, A. Enriched environments, experience-dependent plasticity and disorders of the nervous system. \textit{Nat. Rev. Neurosci.}. \textbf{7}: 697-709 (2006).
\bibitem{Fenton1} Fenton, A., Kao, H., Neymotin, S., Olypher, A., Vayntrub, Y., Lytton, W. \& Ludvig N. Unmasking the CA1 ensemble place code by exposures to small and large environments: more place cells and multiple, irregularly arranged, and expanded place fields in the larger space. \textit{J. Neurosci.} \textbf{28}: 11250-11262 (2008).
\bibitem{Eckert} Eckert, M. \& Abraham, W. Physiological effects of enriched environment exposure and LTP induction in the hippocampus in vivo do not transfer faithfully to in vitro slices. \textit{Learn. \& Mem.} \textbf{17}: 480-484 (2010).

\bibitem{Mizuseki} Mizuseki, K., Sirota, A., Pastalkova, E. \& Buzs\'aki, G. Theta oscillations provide temporal windows for local circuit computation in the entorhinal-hippocampal loop, \textit{Neuron}, \textbf{64}: 267-280 (2009).
\bibitem{Huhn} Huhn, Z., Orb\'an, G., \'Erdi, P. \& Lengyel, M. Theta oscillation-coupled dendritic spiking integrates inputs on a long time scale. \textit{Hippocampus} \textbf{15}: 950-962 (2005).

\bibitem{Ang} Ang, C., Carlson, G. \& Coulter, D. Hippocampal CA1 Circuitry Dynamically Gates Direct Cortical Inputs Preferentially at Theta Frequencies. \textit{J. Neurosci.} \textbf{25}: 9567-9580 (2005).
\bibitem{Maurer} Maurer, A., Cowen, S., Burke, S., Barnes, C. \& McNaughton, B. Organization of hippocampal cell assemblies based on theta phase precession. \textit{Hippocampus} \textbf{16}: 785-794 (2006).

\bibitem{Draguhn} Buzs\'aki, G. \& Draguhn, A. Neuronal Oscillations in Cortical Networks. \textit{Science} \textbf{304}: 1926-1929 (2004).

\bibitem{BuzsakiTheta1} Buzs\'aki, G. Theta oscillations in the hippocampus. \textit{Neuron} \textbf{33}: 325-340 (2002).
\bibitem{Hasselmo} Hasselmo, M., Bodelon, C. \& Wyble, B. A proposed function for hippocampal theta rhythm: separate phases of encoding and retrieval enhance reversal of prior learning. \textit{Neural Computation} \textbf{14}: 793-817 (2002).
\bibitem{Colgin1} Colgin, L. \& Moser, E. Gamma oscillations in the hippocampus. \textit{Physiology} \textbf{25}: 319-329 (2010).
\bibitem{Lisman1} Lisman, J. \& Idiart, M. Storage of 7 +/- 2 short-term memories in oscillatory subcycles. \textit{Science} \textbf{267}: 1512-1515 (1995).

\bibitem{BuzsakiTheta2} Buzs\'aki, G. Theta rhythm of navigation: link between path integration and landmark navigation, episodic and semantic memory. \textit{Hippocampus}, \textbf{15}: 827-840 (2005).
\bibitem{Huxter} Huxter, J., Senior, T., Allen, K. \& Csicsvari, J. Theta phase-specific codes for two-dimensional position, trajectory and heading in the hippocampus. \textit{Nature Neurosci.} \textbf{11}: 587-594 (2008).

\bibitem{Skaggs} Skaggs, W., McNaughton, B., Wilson, M. \& Barnes, C. Theta phase precession in hippocampal neuronal populations and the compression of temporal sequences. \textit{Hippocampus} \textbf{6}: 149-172 (1996).
\bibitem{Jensen3} Jensen, O. \& Lisman, J. Hippocampal CA3 region predicts memory sequences: accounting for the phase precession of place cells. \textit{Learn. \& Mem.} \textbf{3}: 279-287 (1996).
\bibitem{BuzsakiGamma} Buzs\'aki, G. \& Wang, X. Mechanisms of gamma oscillations. \textit{Annual Rev. Neurosci} \textbf{35}: 203-225 (2012).
\bibitem{Colgin2} Colgin, L., Denninger, T., Fyhn, M., Hafting, T., Bonnevie, T., Jensen, O., Moser, M-B. \& Moser, E. Frequency of gamma oscillations routes flow of information in the hippocampus. \textit{Nature} \textbf{462}: 353-357 (2009).

\bibitem{Dragoi} Dragoi, G. \& Buzs\'aki, G. Temporal encoding of place sequences by hippocampal cell assemblies. \textit{Neuron} \textbf{50}: 145-157 (2006).

\bibitem{Jia} Jia, X. \& Kohn, A. Gamma rhythms in the brain. \textit{PLoS Biology} \textbf{9}: e1001045 (2011).
\bibitem{Nikoli} Nikoli, D., Fries, P. \& Singer, W. Gamma oscillations: precise temporal coordination without a metronome. \textit{Trends in cognitive sciences} \textbf{17}: 54-55 (2013).
\bibitem{Lisman3} Lisman, J. The theta/gamma discrete phase code occuring during the hippocampal phase precession may be a more general brain coding scheme. \textit{Hippocampus} \textbf{15}: 913-922 (2005).

\bibitem{Whittington} Whittington, M., Faulkner, H., Doheny, H. \& Traub, R. Neuronal fast oscillations as a target site for psychoactive drugs. \textit{Pharmacology \& Therapeutics} \textbf{86}: 171-190 (2000).
\bibitem{McCracken} McCracken, C. \& Grace, A. Persistent Cocaine-Induced Reversal Learning Deficits Are Associated with Altered Limbic Cortico-Striatal Local Field Potential Synchronization. \textit{J. Neurosci.} \textbf{33}: 17469-17482 (2013).
\bibitem{Dilgen} Dilgen, J., Tompa, T., Saggu, S., Naselaris, T. \& Lavin, A. Optogenetically evoked gamma oscillations are disturbed by cocaine administration. \textit{Frontiers Cell. Neurosci.} \textbf{7}: 213(2013).
\bibitem{Vreugdenhil} Vreugdenhil, M. \& Toescu, E.C. Age-dependent reduction of γ oscillations in the mouse hippocampus in vitro. \textit{Neuroscience} \textbf{132}: 1151-1157 (2005).
\bibitem{Lu} Lu, C., Hamilton, J., Powell, A., Toescu, E. \& Vreugdenhil, M. Effect of ageing on CA3 interneuron sAHP and $\gamma$ oscillations is activity-dependent.\textit{ Neurobiology of aging} \textbf{32}: 956-965 (2011).
\bibitem{Vugt} van Vugt, M., Schulze-Bonhage, A., Litt, B., Brandt, A. \& Kahana, M. Hippocampal Gamma Oscillations Increase with Memory Load. \textit{J. Neurosci.} \textbf{30}: 2694-2699 (2010).
\bibitem{Moretti} Moretti, D., Fracassi, C., Pievani, M., Geroldi, C., Binetti, G., Zanetti, O., Sosta, K., Rossini, P. \& Frisoni, G. Increase of $\theta/\gamma$ ratio is associated with memory impairment. \textit{Clinical Neurophysiology} \textbf{120}: 295-303 (2009).
\bibitem{Lundqvist} Lundqvist, M., Herman, P. \& Lansner, A. Theta and Gamma Power Increases and Alpha/Beta Power Decreases with Memory Load in an Attractor Network Model. \textit{J. Cog. Neurosci.} \textbf{23}: 3008-3020 (2011).
\bibitem{Trimper} Trimper, J., Stefanescu, R. \& Manns, J. Recognition memory and $\theta-\gamma$ interactions in the hippocampus. \textit{Hippocampus} \textbf{24}: 341-353 (2014).
\bibitem{Ulanovsky} Ulanovsky, N. \& Moss, C. Hippocampal cellular and network activity in freely moving echolocating bats. \textit{Nat. Neurosci.} \textbf{10}: 224-233 (2007).
\bibitem{Yartsev} Yartsev, M. \& Ulanovsky, N. Representation of Three-Dimensional Space in the Hippocampus of Flying Bats. \textit{Science} \textbf{340}: 367-372 (2013).

\bibitem{Harris1} Harris, K., Csicsvari, J., Hirase, H., Dragoi, G. \& Buzs\'aki, G. Organization of cell assemblies in the hippocampus, \textit{Nature}: \textbf{424}: 552-556 (2003).
\bibitem{Harris2} Harris, K. Neural signatures of cell assembly organization. \textit{Nat. Rev. Neurosci.} \textbf{6}: 399-407 (2005).
\bibitem{Syntax} Buzs\'aki, G. Neural syntax: cell assemblies, synapsembles, and readers, \textit{Neuron}, \textbf{68}: 362-385 (2010).

\bibitem{Jackson} Jackson, J. \& Redish, A. Network dynamics of hippocampal cell-assemblies resemble multiple spatial maps within single tasks. \textit{Hippocampus} \textbf{17}: 1209-1229 (2007).

\bibitem{ONeill} O'Neill, J., Senior, T., Allen, K., Huxter, J. \& Csicsvari, J. Reactivation of experience-dependent cell assembly patterns in the hippocampus. \textit{Nat. Neurosci.} \textbf{11}: 209-215 (2008).

\bibitem{Burgess} Burgess, N. \& O'Keefe, J. Cognitive graphs, resistive grids, and the hippocampal representation of space. \textit{J. Gen. Physiol.} \textbf{107}: 659-662 (1996).
\bibitem{Muller} Muller, R., Stead, M. \& Pach, J. The hippocampus as a cognitive graph. \textit{J. Gen. Physiol.} \textbf{107}: 663-694 (1996).

\bibitem{Jonsson} Jonsson, J. \textit{Simplicial complexes of graphs}, Springer, New York (2008).

\bibitem{Katz} Katz, Y., Kath, W., Spruston, N. \& Hasselmo, M. Coincidence detection of place and temporal context in a network model of spiking hippocampal neurons. \textit{PLoS Computat. Biol.} \textbf{3}: e234 (2007).
\bibitem{Brette} Brette, R. Computing with Neural Synchrony. \textit{PLoS Comput. Biol.} \textbf{8}: e1002561 (2012).

\bibitem{Konig} K\"{o}nig, P., Engel, A. \& Singer, W. Integrator or coincidence detector? The role of the cortical neuron revisited. \textit{Trends Neurosci.}, \textbf{19}: 130-137 (1996).
\bibitem{Ratte} Ratt\'{e}, S., Lankarany, M., Rho Y-A., Patterson, A. \& Prescott, S. Subthreshold membrane currents confer distinct tuning properties that enable neurons to encode the integral or derivative of their input. \textit{Front. Cell Neurosci.}, \textbf{8}: 452 (2015).
\bibitem{Magee1} Magee, J. Dendritic integration of excitatory synaptic input. \textit{Nat. Rev. Neurosci.} \textbf{1}: 181-190 (2000).
\bibitem{London} London, M. \& H\"ausser, M. Dendritic Computation. \textit{Ann. Rev. Neurosci.} \textbf{28}: 503-532 (2005).
\bibitem{Spruston} Spruston, N. Pyramidal neurons: dendritic structure and synaptic integration. \textit{Nat. Rev. Neurosci.} \textbf{9}: 206-221 (2008).

\bibitem{Lee} Lee, I., Yoganarasimha, D., Rao, G. \& Knierim, J. Comparison of population coherence of place cells in hippocampal subfields CA1 and CA3. \textit{Nature} \textbf{430}: 456-459 (2004).
\bibitem{Jezek} Jezek, K., Henriksen, E., Treves, A., Moser, E. \& Moser, M.-B. Theta-paced flickering between place-cell maps in the hippocampus. \textit{Nature} \textbf{478}: 246-249 (2011).
\bibitem{Geisler} Geisler, C., Diba, K., Pastalkova, E., Mizuseki, K., Royer, S. \& Buzs\'aki, G. Temporal delays among place cells determine the frequency of population theta oscillations in the hippocampus. \textit{Proc. Natl. Acad. Sci.} \textbf{107}: 7957-7962 (2010).
\bibitem{HarrisTheta} Harris, K., Henze, D., Hirase, H., Leinekugel, X., Dragoi, G., Czurk\'o, A. \& Buzs\'aki G. Spike train dynamics predicts theta-related phase precession in hippocampal pyramidal cells. \textit{Nature} \textbf{417}: 738-741 (2002).

\bibitem{Shepherd} Shepherd, G. \textit{The synaptic organization of the brain}, Edn. 5th. Oxford University Press, Oxford; New York (2004).

\bibitem{Neves} Neves, G., Cooke, S., \& Bliss, T. Synaptic plasticity, memory and the hippocampus: a neural network approach to causality. \emph{Nat. Rev. Neurosci.} \textbf{9}: 65-75 (2008).

\bibitem{Barabasi} Barab\'asi, A.-L. \& Albert, R. Emergence of Scaling in Random Networks. \textit{Science} \textbf{286}: 509-512 (1999).
\bibitem{Albert} Albert, R. \& Barab\'asi, A.-L. Statistical mechanics of complex networks. \textit{Reviews of Modern Physics} \textbf{74}: 47-97 (2002).
\bibitem{Li} Li, X., Ouyang, G., Usami, A., Ikegaya, Y. \& Sik, A. Scale-Free Topology of the CA3 Hippocampal Network: A Novel Method to Analyze Functional Neuronal Assemblies. \textit{Biophysical journal} \textbf{98}: 1733-1741 (2010).
\bibitem{Bonifazi} Bonifazi, P., Goldin, M., Picardo, M., Jorquera, I., Cattani, A., Bianconi, G., Represa, A., Ben-Ari, Y. \& Cossart, R. GABAergic Hub Neurons Orchestrate Synchrony in Developing Hippocampal Networks. \textit{Science} \textbf{326}: 1419-1424 (2009).

\bibitem{Selkoe} Selkoe, D. Alzheimer's Disease Is a Synaptic Failure. \textit{Science} \textbf{298}: 789-791 (2002).
\bibitem{Toth} Toth, M., Melentijevic, I., Shah, L., Bhatia, A., Lu, K., Talwar, A., Naji, H., Ibanez-Ventoso, C., Ghose, P., Jevince, A., Xue, J., Herndon, L., Bhanot, G., Rongo, C., Hall, D. \& Driscoll, M. Neurite Sprouting and Synapse Deterioration in the Aging \textit{Caenorhabditis elegans} Nervous System. \textit{J. Neurosci.} \textbf{32}: 8778-8790 (2012).

\bibitem{Vaccarino} Donato, I., Gori M., Pettini M., Petri G., De Nigris S., Franzosi R. \& Vaccarino F. Persistent homology analysis of phase transitions. \textit{Phys. Rev. E} \textbf{93}: 052138 (2016).
\bibitem{Ambjorn} Ambj\o rn, J., Carfora, M. \& Marzuoli, A. \textit{The geometry of dynamical triangulations}, Berlin; New York: Springer (1997).

\bibitem{Wood} Wood, E., Dudchenko, P., Robitsek, R. \& Eichenbaum, H. Hippocampal neurons encode information about different types of memory episodes occurring in the same location. \textit{Neuron} \textbf{27}: 623-633 (2000).
\bibitem{Ginther} Ginther, M., Walsh, D. \& Ramus, S. Hippocampal Neurons Encode Different Episodes in an Overlapping Sequence of Odors Task. \textit{J. Neurosci.} \textbf{31}: 2706-2711 (2011).
\bibitem{Wixted} Wixted, J., Goldinger, S., Squire, L., Kuhn, J., Papesh, M., Smith, K., Treiman, D. \& Steinmetz, P. Coding of episodic memory in the human hippocampus. \textit{Proc. Natl. Acad. Sci.} \textbf{115}: 1093-1098 (2018).

\bibitem{Agster} Agster, K., Fortin N. \& Eichenbaum H. The hippocampus and disambiguation of overlapping sequences. \textit{J Neurosci} \textbf{22}: 5760-5768 (2002). 
\bibitem{Fortin1} Fortin, N., Agster, K. \& Eichenbaum H. Critical role of the hippocampus in memory for sequences of events. \textit{Nat. Neurosci.} \textbf{5}: 458-462 (2002).

\bibitem{EichMem} Eichenbaum, H., Dudchenko, P., Wood, E., Shapiro, M. \& Tanila, H. The hippocampus, memory, and place cells: is it spatial memory or a memory space? \textit{Neuron} \textbf{23}: 209-226 (1999).

\bibitem{Alexandroff} Alexandroff, P. Diskrete R\"aume. \textit{Rec. Math. (Matematicheski Sbornik)} \textbf{2}: 501-518 (1937).

\bibitem{Hopfield} Hopfield, J. Neurodynamics of mental exploration. \textit{Proc. Natl. Acad. Sci.}  \textbf{107}: 1648-1653 (2010).
\bibitem{Issa} Issa, J. \& Zhang, K. Universal conditions for exact path integration in neural systems. \textit{Proc. Natl. Acad. Sci.} \textbf{109}: 6716-6720 (2012).
\bibitem{Dabaghian1} Dabaghian, Y. Maintaining Consistency of Spatial Information in the Hippocampal Network: A Combinatorial Geometry Model. \textit{Neural Comput}: \textbf{28}(6): 1051-1071 (2016).

\bibitem{Buzsaki2} Buzs\'aki, G., Peyrache, A. \& Kubie, J. Emergence of Cognition from Action. \textit{Cold Spring Harb. Symp. Quant. Biol.} \textbf{79}: 41-50 (2014).
\bibitem{Samsonovich} Samsonovich, A. \& McNaughton, B. (1997) Path integration and cognitive mapping in a continuous attractor neural network model. \textit{J Neurosci} \textbf{17}: 5900-5920 (1997).

\bibitem{Nadel} Nadel, L. \& Hardt, O. The spatial brain. \textit{Neuropsychology}, \textbf{18}: 473-476 (2004).

\bibitem{Stong} Stong, R. Finite topological spaces. \textit{Trans. Amer. Math. Soc.} \textbf{123}: 325-340 (1966).
\bibitem{McCord} McCord, M. Singular homology groups and homotopy groups of finite topological spaces. \textit{Duke Math. J.} \textbf{33}: 465-474 (1966).

\bibitem{Cohn} Cohn, A. \& Hazarika, S. Qualitative Spatial Representation and Reasoning: An Overview.\textit{ Fundam Inf} \textbf{46}: 1-29 (2001).
\bibitem{Roeper} Roeper, P. Region-Based Topology. \emph{Journal of Philosophical Logic} \textbf{26}: 251-309 (1997).
\bibitem{Vickers} Vickers, S. \textit{Topology via logic}. Cambridge University Press. (1989).

\end{thebibliography}
\end{document}